\documentclass{iopart}

\input{epsf}

\begin{document}
\jl{6}

\newcommand{\beq}{\begin{equation}}
\newcommand{\eeq}{\end{equation}}
\newcommand{\lab}{\label}
\newcommand{\dd}{{\rm d}}
\newcommand{\ee}{{\rm e}}
\newcommand{\nab}{\nabla\,}
\newcommand{\IR}{{\rm I\!R}}
\newcommand{\blambda}{{\mbox{\boldmath$\lambda$}}}


\letter{Observables and gauge invariance in the theory
of non-linear spacetime perturbations}

\author{Marco Bruni\dag\footnote[3]{E-mail:
bruni@astro.cf.ac.uk} and Sebastiano
Sonego\ddag\footnote[4]{E-mail:
sebastiano.sonego@dic.uniud.it}}
\address{\dag\ Department of Physics and Astronomy,
Cardiff University, Cardiff CF2 3YB, United Kingdom}
\address{\ddag\ Universit\`a di Udine,
DIC -- Via delle Scienze 208, 33100 Udine, Italy}

\date{\today}

\begin{abstract}
 We discuss the issue of observables in general-relativistic
perturbation theory, adopting the view that any observable in general
relativity is represented by a scalar field on spacetime.  In the
context of perturbation theory, an observable is therefore a scalar
field on the perturbed spacetime, and as such is gauge invariant in an
exact sense (to all orders), as one would expect.  However,
perturbations are usually represented by fields on the background
spacetime, and expanded at different orders into contributions that
may or may not be gauge independent.  We show that perturbations of
scalar quantities are observable if they are first order
gauge-invariant, even if they are gauge dependent at higher
order. Gauge invariance to first order plays therefore an important
conceptual role in the theory, for it selects the perturbations with
direct physical meaning from those having only a mathematical status.
The so-called ``gauge problem'', and the relationship between measured
fluctuations and gauge dependent perturbations that are computed in
the theory are also clarified.
\end{abstract}

\vspace{1.5truecm}

The issue of what is observable in general relativity is
still a controversial one\cite{rovelli}. Here we want to
address a related problem, which is fundamental for
practical purposes: which {\em perturbations\/} are
observable in general relativity? Clearly, the answer to
this question is largely conditioned on one's attitude
towards the more basic issue mentioned above. However, we
believe that this question deserves an analysis of its own,
given the peculiar nature of gauge issues in the context of
relativistic perturbation theory, and the fact that the
comparison of Einstein's theory with observations is almost
entirely based on approximation methods (see e.g.\
\cite{bi:schutz} for a discussion of this point).

In the following, we adopt the view that an observable
quantity in general relativity\footnote{Our discussion is
however valid in any  spacetime theory.} is simply
represented by a scalar field on spacetime.  This definition
may seem naive, but it corresponds exactly to what most
physicists have in mind when thinking about observable
quantities.  At a more sophisticated level, it can be argued
that such a notion of observables is practically viable even
if one considers the  issues related to the invariance
of general relativity under spacetime diffeomorphisms.

In the relativistic theory of perturbations one is always
dealing with two spacetimes, the physical (perturbed) one,
and an idealised (unperturbed) background. In this context,
given the above definition, an observable is a scalar on the
perturbed spacetime. Thus, physical quantities such as,
e.g., the energy density in a cosmological model, trivially
satisfy our definition of observables. However, in
perturbation theory one formally decomposes quantities of
the physical space into the sum of a background quantity and
a perturbation. The question arises therefore, whether
perturbations themselves can be regarded as observables.
Indeed, it is certainly useful to express the perturbative
formalism directly in terms of variables (the perturbations)
that are observable quantities, or at least to have clear in
mind how to relate mathematical variables with physically
meaningful ones. Given the recent development of second
order perturbation theory in cosmology (see \cite{mmb,mps} and
references therein) and black hole physics (see
\cite{price,CL} and references therein), the issue becomes
even more important, because of the further complications
that arise when non-linearities are considered.

As we shall discuss shortly, the perturbations are
represented by fields on the background, and expanded at
different orders into contributions that may or may not be
gauge independent. On the other hand, one would expect that
observables are described by gauge-invariant quantities.
Recently, explicit calculations have shown that fields which
are usually regarded as representing observables are not
gauge-invariant at second order\cite{mmb,mps,CL}, although it
has long been known that they are gauge-invariant at first
order (see e.g.\ \cite{teukolski,SW,EB}). In this letter we
point out that the perturbation of a scalar describes an
observable if and only if its representation on the
background is gauge-invariant at first order, even when it
is gauge dependent at higher orders. We shall also discuss
how this seemingly paradoxical result corresponds to have
observable perturbations that are gauge-invariant in the
exact sense (i.e., at all orders) in the physical spacetime,
as expected. Finally, we shall briefly comment on how a
first order gauge dependent perturbation can acquire
physical meaning in a specific gauge.

Let us begin by reviewing some general ideas about the
perturbative approach in general
relativity.\footnote{Hereafter, we shall adopt the
conventions and notations of references \cite{bmms,sb}. See
also reference \cite{ycm} for the more general mathematical
definitions.} Suppose that the physical and the background
spacetimes are represented by the Lorentzian manifolds
$({\cal M},g)$ and $({\cal M}_0,g_0)$, respectively (as
manifolds, ${\cal M}_0={\cal M}$, but it is nevertheless
convenient to label them differently). The perturbation of a
quantity should obviously be defined as the difference
between the values that the quantity takes in $\cal M$ and
${\cal M}_0$, evaluated at points which correspond to the
same physical event. However, there is nothing intrinsic to
$({\cal M},g)$ and $({\cal M}_0,g_0)$ that allows us to
establish a one-to-one correspondence between the two
manifolds. This follows directly from the fact that no
spacetime structure is assigned {\em a priori\/} in general
relativity, contrarily to what happens, e.g., in the
Newtonian theory, where ${\cal M}_0$ and $\cal M$ are simply
identified, thus making possible a straightforward
formulation of Eulerian perturbation theory \cite{stewart}
(see, however, \cite{lagr} and references therein for a
Lagrangian formulation in Newtonian cosmology). Hence, in
general relativity, points of $\cal M$ are unrelated with
points of ${\cal M}_0$, and if we want to compare a physical
quantity in the two spacetimes, we need an additional
prescription about the pairwise identification of points
between $\cal M$ and ${\cal M}_0$. Mathematically, this
corresponds to the assignment of a diffeomorphism
$\varphi:{\cal M}_0\to {\cal M}$, sometimes called a {\em
point identification map\/} \cite{SW}. Such a diffeomorphism
can be given directly as a mapping between ${\cal M}_0$ and
$\cal M$ or --- perhaps more commonly --- by choosing charts
$X$ in ${\cal M}_0$ and $Y$ in $\cal M$ (with coordinates
$\{x^\mu\}$ and $\{y^\mu\}$, respectively), and identifying
points having the same value of the coordinates
\cite{bi:schutz,bardeen}, so that $\varphi$ is implicitly
defined through the relation $x^\mu(p)=y^\mu(\varphi(p))$,
$\forall p\in{\cal M}_0$. In any case, the point
identification map is completely arbitrary; this freedom is
peculiar of general relativistic perturbation theory, and
has no counterpart in theories that are formulated on a
fixed background, where no ambiguity arises in comparing
fields. Following Sachs \cite[p 556]{sachs}, one may refer
to it as gauge freedom ``of the second kind'', in order to
distinguish it from the usual gauge freedom of general
relativity. However, in the following we shall never
consider ordinary gauge transformations of the perturbed and
background spacetimes, and we shall therefore use the term
``gauge'' as synonym of ``gauge of the second kind''.

Once a gauge choice has been made (i.e., a point
identification map $\varphi$ has been assigned),
perturbations can be defined unambiguously. Let $T$ be a
tensor field on $\cal M$. If $T_0$ is the background tensor
field corresponding to $T$, the total {\em perturbation\/}
of $T$ is simply given by $\Delta^\varphi T:=T-\varphi_*
T_0$ \cite{bmms,sb}. By definition, $\Delta^\varphi T$ is a
field on $\cal M$. On the other hand, the aim of  perturbation
theory in general relativity is to {\em construct\/},
through an iterative scheme, the geometry on $\cal M$ (see
e.g. \cite {bi:waldbook}). To this purpose, it is customary
to work with fields on ${\cal M}_0$ (see \cite{bmms,sb} and
references therein). Thus, to start with, the {\em
representation\/} on ${\cal M}_0$ of an arbitrary tensor
field $T$ is defined as the pull-back $\varphi^* T$. Then,
the representation on the background of the perturbation is
$\Delta_0^\varphi T:=\varphi^*\Delta^\varphi T=\varphi^*
T-T_0$. These are the background fields that are
Taylor-expanded to obtain the contributions at different
orders that are used in the above mentioned iteration
scheme.

We would like to point out here that, for practical
purposes, it is often very convenient to use, on ${\cal
M}_0$ and $\cal M$, coordinates $\{x^\mu\}$ and $\{y^\mu\}$
``adapted'' to $\varphi$, such that
$x^\mu(p)=y^\mu(\varphi(p))$, $\forall p\in{\cal M}_0$. In
this way the components of $T$ at the point
$\varphi(p)\in{\cal M}$ coincide with those of $\varphi^* T$
at $p\in{\cal M}_0$ (see figure 1 for a pictorial
explanation in the case of a vector). Obviously, the same is
true for $\varphi_* T_0$ and $T_0$, so that one has, for a
tensor of type $(r,s)$,
\begin{eqnarray} \fl
{\left(\Delta^\varphi T\right)^{\mu_1\ldots\mu_r}}_{\nu_1
\ldots\nu_s}(x)=
{\left(\Delta_0^\varphi T\right)^{\mu_1\ldots\mu_r}}_{\nu_1
\ldots\nu_s}(x)=
{T^{\mu_1\ldots\mu_r}}_{\nu_1\ldots\nu_s}(x)-
{{T_0}^{\mu_1\ldots\mu_r}}_{\nu_1\ldots\nu_s}(x)\;,\nonumber
\\ \lab{chart}\end{eqnarray}
where $x=X(p)=Y(\varphi(p))$. This choice is however rather
confusing from a conceptual point of view because, once the
above identification has been made, it is hard to
distinguish between $\Delta^\varphi T$ and $\Delta^\varphi_0
T$. The point is that, as we shall show, $\Delta^\varphi T$
may be gauge-invariant in an exact sense (at all orders) and
thus, if $T$ is a scalar, it may correspond to an
observable, even when $\Delta^\varphi_0 T$ is
gauge-invariant only at first order.

Under a gauge transformation $\varphi\to\psi$, where $\psi$
is another point identification map, the representation on
${\cal M}_0$ of tensor fields defined on $\cal M$ changes as
under the action of a diffeomorphism.  This can easily be
seen by noticing that a point $p\in{\cal M}$ corresponds, in
two gauges $\varphi$ and $\psi$, to the points
$\varphi^{-1}(p)$ and $\psi^{-1}(p)$ in ${\cal M}_0$.
Defining a map $\Phi:{\cal M}_0\to {\cal M}_0$ as
$\Phi:=\varphi^{-1}\circ\psi$, we have $\varphi^{-1}(p)=
\Phi(\psi^{-1}(p))$.  Then the two representations on
${\cal M}_0$  of a tensor $T$ of ${\cal M}$
are related by
 $\psi^* T=
\left(\varphi\circ\Phi\right)^*T=
\Phi^*\left(\varphi^* T\right)$.
It follows that the gauge transformation between the two
representations of the perturbation of $T$ in the two gauges
is:
\beq
\Delta^\psi_0 T=\Phi^* \Delta^\varphi_0 T+\Phi^*T_0 -T_0
=\Phi^*\left(\varphi^*T\right) -T_0.
\eeq
We want to point out here that from the first equality in
this equation it would seem that a fundamental fact about
gauge transformations is that $\Phi^*$ acts on $T_0$.
However, it is clear from the second equality that this is
not the case, as all is really needed is the action of
$\Phi^*$ on the representation on ${\cal M}_0$ of the field
$T$ of the physical spacetime. In other words, it is only
the fact that we insist to look at the gauge transformation
of $\Delta^\varphi_0 T$ as a whole that brings about the
$\Phi^*T_0$ term in the first equality. In this sense, it
would be an improper extrapolation to say that ``a gauge
transformation is equivalent to a diffeomorphism of the
background''.

\begin{figure}
\centerline{\epsfysize=7cm \epsfbox{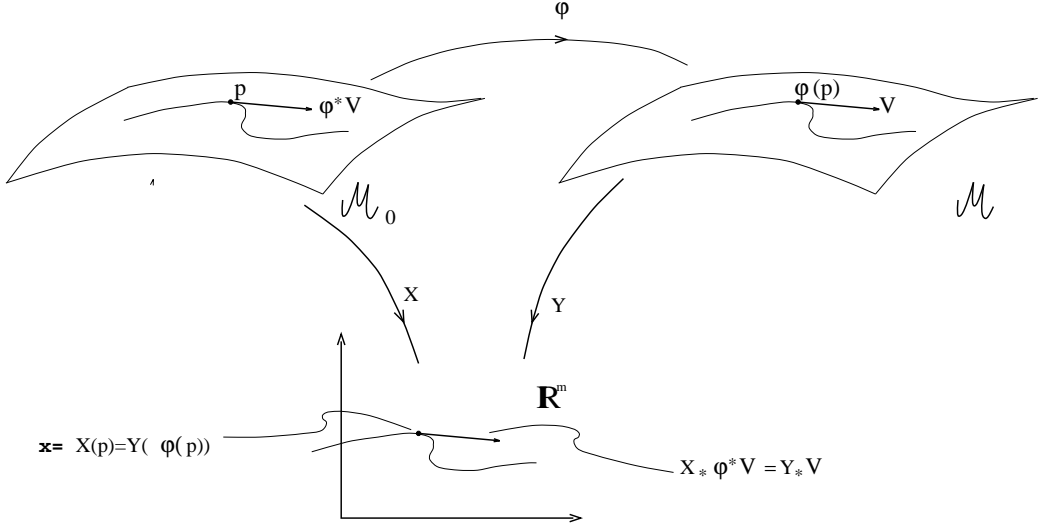}}
\caption{By choosing the coordinates on ${\cal M}_0$ and
$\cal M$ in such a way that $y^\mu\circ\varphi=x^\mu$, a
curve in ${\cal M}_0$ and its $\varphi$-transformed in $\cal
M$ have the same representation in $\IR^m$.  Therefore, the
components of the tangent vectors $V$ and $\varphi^* V$ at
the points $\varphi(p)$ and $p$ are the same:
$\left(\varphi^*V\right)^\mu(x)=
\left(\varphi^*V\right)(x^\mu)|_p=
V\left(x^\mu\circ\varphi^{-1}\right)|_{\varphi(p)}=
V(y^\mu)|_{\varphi(p)}= V^\mu (x)$.}
\end{figure}

It is useful to classify fields on $\cal M$, according to
whether they are {\em intrinsically gauge independent\/}
(IGI) or {\em intrinsically gauge dependent\/} (IGD). We say
that a field is IGI iff its value at any point of $\cal M$
does not depend on the gauge choice; otherwise, we say that
it is IGD. As obvious examples of IGI and IGD quantities we
mention, respectively, a tensor field $T$ defined on $\cal
M$, and the push-forward $\varphi_* T_0$ on $\cal M$ of a
non-trivial\footnote{Hereafter by a trivial tensor we mean
one that is either vanishing or a constant multiple of the
identity \cite{SW}.} tensor field $T_0$ defined on ${\cal M}_0$. It
follows that, given our identification of observables with
scalar functions on $\cal M$ and the arbitrariness in the
choice of the gauge $\varphi$, a scalar describes an
observable only if it is IGI. On the other hand, considering
the representation on ${\cal M}_0$ of the perturbations, one
is led to define a corresponding idea in the background,
saying that a quantity on ${\cal M}_0$ is {\em
identification gauge-invariant\/} (i.g.i.) iff its value at
any point of ${\cal M}_0$ does not depend on the gauge
choice \cite{SW}. An example of i.g.i.\ quantity is a tensor
field $T_0$ defined on ${\cal M}_0$, while the pull-back
$\varphi^\ast T$ of a tensor field $T$ defined on $\cal M$
is not i.g.i.\ unless $T$ is trivial. It is then clear that
the representation on the background of an IGI quantity is
not i.g.i.\ in general, but this is totally irrelevant as
far as the issue of observability is concerned, because
measurements are always performed in the physical spacetime
$\cal M$, whereas the background ${\cal M}_0$ has merely the
status of a useful mathematical artifice.

Let us now turn our attention to perturbations, asking
whether they are IGI or IGD, and how this relates to the
gauge dependence of their representation on ${\cal M}_0$. In
order to answer these questions, we consider a gauge
transformation $\varphi\to\psi$. Correspondingly,
perturbations transform as $\Delta^\varphi T\to\Delta^\psi
T$, where
\beq
\Delta^\psi
T=\Delta^\varphi T+\left(\varphi_* T_0-\psi_* T_0\right)\;.
\lab{transfM}
\eeq
Similarly, their representations on the background
transform as $\Delta_0^\varphi T\to\Delta_0^\psi T$, with
\beq
\Delta_0^\psi T=\Delta_0^\varphi T+\left(\psi^* T-\varphi^*
T\right)\;.  \lab{transfM0}
\eeq
Therefore, in general, both the perturbations on $\cal M$
and their representations on ${\cal M}_0$ change under the
action of a gauge transformation. This gauge dependence of
perturbations does not appear in theories that admit a
canonical identification between $\cal M$ and ${\cal M}_0$,
and is due to the arbitrariness in the choice of a point
identification map, which is additional to the usual gauge
freedom of general relativity. In a sense, general
relativistic perturbations ``have gauge freedom of their
own'', i.e.\ the freedom ``of the second kind"\cite{sachs}
mentioned before, even when the full quantities have not. We
have to remind at this point that, once the Taylor expansion
into different order contributions has been carried out, it
turns out that a perturbation on ${\cal M}_0$ may be
gauge-invariant at first order, and not at higher orders
\cite{bmms,sb}. Now, it follows from (\ref{transfM}) that
the perturbation is IGI iff $\varphi_* T_0=\psi_* T_0$,
$\forall\varphi,\psi$. This can be rewritten as $T_0=\Phi^*
T_0$, where $\Phi:=\varphi^{-1}\circ\psi:{\cal M}_0\to {\cal
M}_0$, and is satisfied only if $T_0$ is trivial. But this
is precisely the condition for first order i.g.i.\ derived
by Stewart and Walker \cite{SW}. Thus, the perturbation of
$T$ is IGI iff its representation on the background is
gauge-invariant to first order. For the particular case in
which $T$ is a scalar physical quantity, we obtain the main
result of this letter: the perturbation of $T$ is observable
iff its representation on ${\cal M}_0$ is first order
i.g.i., even when it is gauge dependent to higher orders.

This result may sound trivial, but only because we have approached the
question of observability of perturbations from the side of the
physical spacetime $\cal M$. Focusing attention only on perturbations
as fields on ${\cal M}_0$ rather than on $\cal M$, as it is usually
done in the iterative scheme, the notion of IGI quantities does not
naturally arise, and one would ask instead whether the representation
of a perturbation is i.g.i., i.e., whether $\Delta_0^\varphi
T=\Delta_0^\psi T$, $\forall\varphi,\psi$ identification maps
\cite{SW}. Because of (\ref{transfM0}), this condition is equivalent
to the requirement that $\varphi^* T=\psi^* T$, $\forall\varphi,
\psi$, which is satisfied iff $T$ is trivial. Thus, the perturbation
of a quantity is IGI iff the quantity itself is trivial in the
background ${\cal M}_0$, while its representation is i.g.i.\ iff the
quantity is trivial on $\cal M$. The important point is that the
physically interesting condition is not i.g.i., but IGI, as one can
clearly see by considering the case in which $T$ is a scalar. The
requirement that the perturbation of $T$ be i.g.i.\ amounts to saying
that $T$ must be a constant on $\cal M$, which is too strong a
constraint to be fulfilled by most observables of physical
interest. On the contrary, IGI requires only that $T_0$ be
constant. As we have already pointed out, the physical content of the
theory resides in the quantities on $\cal M$, not in their
representations on ${\cal M}_0$. From the physical point of view,
there is nothing bad in having a gauge dependent pull-back on ${\cal
M}_0$, provided that the quantity on $\cal M$ be uniquely defined. 
Of course, to the first order in a perturbative
approach, the perturbation of a quantity is IGI iff its representation
is i.g.i.

The gauge dependence of perturbations has often been
referred to in the past as the {\em gauge problem\/}
\cite{EB,bardeen}.\footnote{A secondary issue that follows
from this gauge dependence is that of gauge modes. The
latter arise when the prescription of point identification
has not been given, or when it is not sharp enough to select
a single $\varphi: {\cal M}_0\to {\cal M}$, and defines
instead a {\em class\/} of diffeomorphisms \cite{ks}. Of
course, for gauge-invariant perturbations both problems
disappear.} It is now clear that, for IGI perturbations,
such gauge dependence is only an artifact of focusing
attention on the representation, and there is no real
problem in the physical spacetime $\cal M$. The situation is
different for IGD perturbations. Their very definition
requires a gauge to be defined: this {\em is\/} a
difficulty, because then these quantities cannot directly
represent observables. However, especially in cosmology, it
is often the case that astronomers {\em do\/} measure
quantities that they call perturbations, or fluctuations,
which they define with respect to {\em an averaged
quantity\/}. It is obvious that such fluctuations must be
IGI, as their very definition does not involve any
background or gauge choice. A paradigmatic example is that
of the Cosmic Microwave Background (CMB) temperature
anisotropy (see \cite{gl1,maartens} and references therein).
We are then facing a paradox: on the one hand theoretical
perturbations that are commonly considered are
 IGD (see e.g. \cite{EB}), whereas their
observational counterpart are IGI.  It is then necessary to
understand what is the relationship between them.

The resolution of the problem lies in recognising the
misleading identification that is often made between
averages in the physical spacetime and background
quantities. Supposing that an averaging procedure has been
adopted,\footnote{Such an average can be performed on data taken on
suitable subspaces of the
whole spacetime.  A
particularly interesting case is that of averages on the sky
made by each observer at his spacetime point. Obviously, in
general a sky-averaged quantity is point-dependent, an
important fact for the following discussion.} the measured
fluctuation of a quantity $T$ is defined by the observer
simply as $\Delta T:=T-\langle T\rangle$. Clearly, such a
definition is valid in any spacetime and, by itself, has no
relation whatsoever with the adoption of a background model
and/or of a perturbative formalism (of which the observer
can be happily totally unaware). However, if we want to
consider this spacetime from a perturbative point of view,
then such a relation is easily established by rewriting
$\Delta T$ as
\beq
\Delta T=T- \langle T\rangle=\left(T-\varphi_*T_0\right)-
\left(\langle T \rangle-\varphi_*T_0\right)=\Delta^\varphi
T-\Delta^\varphi \langle T \rangle\;,
\eeq
where the same background quantity $T_0$ has been used for
both $T$ and $\langle T\rangle$.  Thus, the theoretically
computed perturbation, $\Delta^\varphi T$, differs from the
measured one, $\Delta T$, by the term $\Delta^\varphi
\langle T \rangle$. The apparent paradox outlined above
arises therefore when $\Delta^\varphi T$ is identified with
$\Delta T$, which happens if one thinks of the average
$\langle T \rangle$ as the push-forward $\varphi_* T_0$ of
some $T_0$ defined in the background spacetime. This is
generally wrong, as one can see by considering the example
of the CMB anisotropy, where $\langle T\rangle$ is obtained
by an angular average \cite{gl1,maartens} and depends
therefore on the spatial position as well as on time, while
$T_0$ can depend only on time, being the temperature in a
spatially homogeneous cosmological model. The only case in
which the identification is meaningful is when $\langle
T\rangle$ and $T_0$ are constant on spacetime. Indeed,
$\langle T \rangle$ is IGI by definition, while in general
$\varphi_* T_0$ is IGD. If we want $\langle T\rangle=
\varphi_* T_0$, it is clear that $\varphi_* T_0$ must also
be IGI, which can be the case only if $T_0$ is a constant.
Instead, of particular interest in cosmology is the case
when $\langle T \rangle_0= T_0$, i.e.\ when we consider a
sky or spatial average, so that $\Delta T$ has a vanishing
background value, and therefore is IGI (and, of course,
first order i.g.i.).

Scalars that deserve specific attention are those built by
projecting a tensor over a tetrad: a well-known example is
that of the Weyl scalars in the Newman-Penrose (NP)
formalism, which are often used in the study of black hole
perturbations \cite{CL,teukolski,SW}. Let us consider, in
order to fix the ideas, the NP scalar
$\Psi_4:=C(l,\overline{m},l,\overline{m})$, where $C$ is the
Weyl tensor and $\{ l,n,m,\overline{m}\}$ is the usual NP
null tetrad. The perturbation of $\Psi_4$ is
\begin{equation}
\Delta^\varphi\Psi_4=\Psi_4-\varphi_\ast\Psi_{4,0}\;,
\lab{pertweyl}\end{equation}
where
$\Psi_{4,0}=C_0(l_0,\overline{m}_0,l_0,\overline{m}_0)$ is
constructed only from quantities defined on ${\cal M}_0$. It
may seem that $\Delta^\varphi\Psi_4$ contains more
arbitrariness than the perturbations we have considered so
far, because its definition requires not only to choose a
gauge $\varphi$, but also to specify a tetrad $\{
l_0,n_0,m_0,\overline{m}_0\}$ in the background and another
one, $\{ l,n,m,\overline{m}\}$, in the physical space.
Actually, the two tetrads are not independent, because if we
require that $\Psi_4\to\Psi_{4,0}$ in the limit of vanishing
perturbations, we must have, e.g., $l=\varphi_\ast
l_0+\Delta^\varphi l$, so that the perturbed tetrad can be
constructed iteratively from the unperturbed one, by
imposing that it be null with respect to the perturbed
metric, order by order. Nevertheless, we are still left with
a dependence of $\Delta^\varphi\Psi_4$ on the background
tetrad. However, it is easy to understand that this tetrad
dependence is not a problem as far as observability is
concerned. Indeed, changing the tetrad from $\{
l_0,n_0,m_0,\overline{m}_0\}$ to $\{
l'_0,n'_0,m'_0,\overline{m}'_0\}$, say, $\Psi_4$ will change
to another scalar $\Psi'_4$ with a {\em different\/}
physical/geometrical interpretation. Tetrad dependence
corresponds thus to the possibility of constructing
different measurable quantities starting from the Weyl
tensor, and has not to be regarded, conceptually, on the
same footing of the gauge dependence that we have considered
in the rest of this letter. Also, in practice one is guided
by the Petrov algebraic type of the background in choosing on
it a specific tetrad, such that some of the Weyl scalars
vanish and thus are first order i.g.i. For example,
Schwarzchild and Kerr spacetimes are Petrov type D, and the
tetrad can be aligned with the principal null directions, so that 
 only $\Psi_2\not =0$, which
results in $\Psi_0$ and $\Psi_4$ describing gravitational
wave perturbations in a gauge-invariant way
\cite{CL,teukolski,SW}. At second order, the construction of
a gauge {\em and\/} tetrad invariant NP perturbative
formalism is computationally very useful, e.g.\ in order to
compare results with those of a fully numerical treatment.
For Kerr, this can be achieved \cite{CL} by implementing a
procedure that strongly reminds the one used at first order
in cosmology \cite{bardeen}. One can construct infinitely
many of such gauge-invariant quantities, but again physical
meaning guides the choice: in \cite{CL} the chosen
second-order i.g.i.\ quantity is the one reducing to
$\Psi_4$ in an asymptotically flat gauge.

Finally, one may wonder what gauge dependent variables,
which are commonly used in perturbation theory, have to do
with observables. The answer to this question is obvious:
provided that calculations are free from gauge modes, even a
first order gauge dependent perturbation acquires physical
meaning in a specific gauge when, in that gauge, it can be
identified with a gauge-invariant quantity.  This is the
case, for example, of the density perturbation $\delta
\rho/\rho$ in cosmology: its value in the comoving gauge
coincides with the value in that gauge of one of Bardeen's
gauge-invariant variables \cite{bardeen}, and in turn the
latter is the first order expansion of a covariantly defined
density gradient \cite{bde}. It is our opinion that there is
more appeal in working directly with observable variables
which are automatically IGI, such as the fluctuation $\Delta
T$ considered above. However, it may often be the case that
one is not able to find a complete set of such quantities,
so that one will either resort to gauge-dependent variables,
or to appropriate combinations of gauge-dependent quantities
that are gauge-invariant at the desidered order
\cite{CL,bardeen}.


\ack
We thank Manuela Campanelli and Carlos Lousto for valuable
discussions. We are grateful for hospitality to the
Department of Astronomy and Astrophysics of Chalmers
University, G\"oteborg, and to the Astrophysics Sector of
SISSA, Trieste, where part of this work was carried out.


\end{document}